\documentclass[10pt, conference]{IEEEtran}
\usepackage{subfigure}
\usepackage{setspace}
\usepackage{amsmath}
\usepackage{amssymb}
\usepackage{amsfonts}
\usepackage{amscd}
\usepackage{mathrsfs}
\usepackage[final]{graphicx}
\usepackage{graphicx}
\usepackage{psfrag}
\usepackage{color}
\usepackage{url}

\newtheorem{definition}{Definition}

\begin{document}

\title{A Low-decoding-complexity, Large coding Gain, Full-rate, Full-diversity STBC for  $4\times2$ MIMO System}

\author{
\authorblockN{K. Pavan Srinath}
\authorblockA{Dept of ECE, Indian Institute of science \\
Bangalore 560012, India\\
Email:pavan@ece.iisc.ernet.in\\
}
\and
\authorblockN{B. Sundar Rajan}
\authorblockA{Dept of ECE, Indian Institute of science \\
Bangalore 560012, India\\
Email:bsrajan@ece.iisc.ernet.in\\
}
}

\maketitle
\begin{abstract}
This paper proposes a low decoding complexity, full-diversity and full-rate space-time block code (STBC) for 4 transmit and 2 receive ($4\times 2$) multiple-input multiple-output (MIMO) systems. For such systems, the best code known is the DjABBA code and recently, Biglieri, Hong and Viterbo have proposed another STBC (BHV code) which has lower decoding complexity than DjABBA but does not have full-diversity like the DjABBA code. The code proposed in this paper has the same decoding complexity as the BHV code for square QAM constellations but has full-diversity as well. Compared to the best code in the DjABBA family of codes, our code has lower decoding complexity, a better coding gain and hence a better error performance as well. Simulation results confirming these are presented.
\end{abstract}

\section{INTRODUCTION}
Multiple-input multiple-output (MIMO) transmission has been of special interest in wireless communication for the past one decade. The Alamouti code \cite{SMA} for two transmit antennas, due to its orthogonality properties, allows a low complexity maximum-likelihood (ML) decoder. This scheme paved the way for generalized orthogonal STBCs \cite{TJC}. Such codes allow the transmitted symbols to be decoupled from one another and single-symbol ML decoding is achieved over quasi-static Rayleigh fading channels. Another aspect of these codes is that they achieve the maximum diversity gain for  any number of transmit and receive antennas and for any arbitrary complex constellations. Unfortunately, for more than two antennas, rate 1 codes  cannot be constructed using orthogonal designs. 

With a view of increasing the transmission rate, quasi-orthogonal designs (QODs) were proposed in \cite{JH}. However, these codes come at the price of a smaller diversity gain and are also double symbol decodable for 4 antennas. As an improvement, Coordinate interleaved orthogonal designs (CIODs) were proposed \cite{KhR}. These codes have the same transmission rate as QODs but additionally enjoy full diversity while being single symbol decodable for certain complex constellations. But none of the above class of codes is  \textit{full-rate}, where an STBC is said to be of full-rate if its rate in complex symbols per channel use is equal to the minimum of the number of transmit and the receive antennas. 

Full-rate, full-diversity STBCs are of prime importance in systems like WIMAX. Low-decoding complexity, full-rate STBCs have been proposed in \cite{PGA} and \cite{SS} for $2\times 2$ and in \cite{BHV} for $4\times 2$ MIMO systems. These codes allow a simplified ML decoding when compared with codes from division algebras \cite{BRV},\cite{SRS} which are not amenable for low decoding complexity though they offer full-rate. The fast decodable code proposed in \cite{BHV} for $4\times 2$ systems,  which we call the BHV code, outperforms the best known DjABBA code only at low SNRs while allowing a reduction in the ML decoding complexity. The BHV code does not have full-diversity as it is based on the quasi orthogonal design for 4 antennas, when all the symbols are take values from one constellation.

In this paper, we propose a new STBC for $4\times2$ MIMO transmission. Our code is based on the Coordinate Interleaved Orthogonal Designs (CIODs) proposed in  \cite{KhR} (defined in Section \ref{sec3}). The major contributions of this paper can be summarized as follows:
\begin{itemize}
 \item Our code has a decoding complexity of the order of $M^5$, for all complex constellations, where $M$ is the size of the signal constellation, whereas the DjABBA code has the corresponding complexity of order $M^7$ and the BHV code has order $M^6,$ ($M^5$ for square QAM constellations - though this has not been claimed in \cite{BHV}). 

\item Our code has a better CER ( Codeword error rate ) performance than the best code in the DjABBA family due to a higher coding gain for QAM constellations.  
\item Our code outperforms the BHV code for QAM constellations due to its higher diversity gain.
\item Combining the above, it can be seen that when QAM constellations are used, our code is the best among all known codes for $4 \times 2$ systems.
\end{itemize}

The remaining content of the paper is organized as follows: In Section \ref{sec2}, the system model and the code design criteria is given. The proposed STBC and its decoding complexity are discussed in Section \ref{sec3}. In Section \ref{sec4}, the decoding scheme for the proposed STBC using sphere decoding is discussed. In Section \ref{sec5}, simulation results are presented to illustrate the comparisons with best known codes. Concluding remarks constitute Section \ref{sec6} .

\textit{Notations:} Let $X$ be a complex matrix. Then $X^T$, $X^{H}$ and $det\left[X\right]$ denote the transpose, Hermitian and the determinant of $X$ respectively. $\mathcal{R}\left(s\right)$ and $\mathcal{I}\left(s\right)$ denote the real and imaginary parts of a complex number $s$, respectively, and  $j$ represents $\sqrt{-1}.$ The set of all real and complex numbers are denoted by $\mathbb{R}$ and $\mathbb{C}$, respectively. $\Vert.\Vert_F$ and $\Vert.\Vert$ denote the Frobenius norm and the vector norm, respectively and $tr\left[.\right] $ denotes the trace operation. For a matrix $X,$ the vector obtained by columnwise concatenation one below the other is denoted by ${vec\left(X\right)}.$  The Kronecker product is denoted by $\otimes$ and $I_T$ denotes the $T\times T$ identity matrix. Given a complex vector $\textbf{x} = \left[ x_1, x_2, \cdots, x_n \right]^T,$ the vector $\tilde{\textbf{x}}$ is defined as
\begin{equation*}
\tilde{\textbf{x}} \triangleq \left[ \mathcal{R}\left(x_1\right), \mathcal{I}\left(x_1\right), \cdots, \mathcal{I}\left(x_n\right)\right ]^T 
\end{equation*}
and for a complex number $s$, the matrix $\check{{\bf s}}$ operator is defined by
\begin{equation*}
\check{\textbf{s}} \triangleq \left[\begin{array}{cc}
\mathcal{R}\left(s\right) & -\mathcal{I}\left(s\right)\\
\mathcal{I}\left(s\right) &  \mathcal{R}\left(s\right)\\
\end{array}\right]
\end{equation*}
The $\check{\left(.\right)}$ operator can be similarly applied to $n\times n$ matrix by applying it to all the entries.

\section{SYSTEM MODEL}
\label{sec2}
We consider Rayleigh quasi-static flat-fading MIMO channel with full channel state information (CSI) at the receiver but not at the transmitter. For $4\times2$ MIMO transmission, we have
\begin{equation}\label{Y}
\textbf{Y = HS + N}
\end{equation} 

\noindent where $\textbf{S} \in \mathbb{C}^{4\times4}$ is the codeword matrix, transmitted over 4 channel uses, $\textbf{N} \in \mathbb{C}^{2\times4}$ is a complex white Gaussian noise matrix with i.i.d entries $\sim
\mathcal{N}_{\mathbb{C}}\left(0,N_{0}\right)$ and $\textbf{H} \in \mathbb{C}^{2\times4}$ is the channel matrix with the entries assumed to be i.i.d circularly symmetric Gaussian random variables $\sim \mathcal{N}_\mathbb{C}\left(0,1\right)$. $\textbf{Y} \in \mathbb{C}^{2\times4}$ is the received matrix

\begin{definition}\label{def1}$\left(\textbf{Code rate}\right)$ If there are $k$ independent information symbols in the codeword which are transmitted over $T$ channel uses, then, for an  $n_t\times n_r$ MIMO system, the code rate is defined as $k/T$ symbols per channel use. For instance, for the Alamouti code $k=2$ and $T=2.$ If $k = n_{min}T$, where $n_{min} = min\left(n_t,n_r\right)$, then the STBC is said to have $full$ $rate$.
\end{definition}

\noindent Considering ML decoding, the decoding metric that is to be minimized over all possible values of codewords $\textbf{S}$ is given by
 \begin{equation}
\label{ML}
 \textbf{M}\left(\textbf{S}\right) = \Vert \textbf{Y} - \textbf{HS} \Vert_F^2 
 \end{equation}

\begin{definition}\label{def2}$\left(\textbf{Decoding complexity}\right)$ 
The ML decoding complexity can be measured by the minimum number of values of $\textbf{M}\left(\textbf{S}\right)$ that are needed to be computed in minimizing the decoding metric.
\end{definition}
\begin{definition}\label{def3}$\left(\textbf{Generator matrix}\right)$ For any STBC $\textbf{S}$ that encodes $k$ information symbols, the $generator$ matrix $\textbf{G}$ is defined by the following equation
\begin{equation}
\widetilde{vec\left(\textbf{S}\right)} = \textbf{G} \tilde{\textbf{s}}.
\end{equation}
\noindent where $\textbf{s} \triangleq \left[ s_1, s_2,\cdots,s_k \right]^T$ is the information symbol vector \cite{BHV}.
\end{definition}

Code design is based on the analysis of pairwise error probability (PEP) given by $P(\textbf{X} \rightarrow \hat{\textbf{X}})$, which is the probability that a transmitted codeword $\textbf{X}$ is detected as $\hat{\textbf{X}}$. The goal is to minimize the error probability, which is upper bounded by the following union bound.
\begin{equation}
P_{e} \leq \frac{1}{M^{k}}\sum_{\textbf{X}} \sum_{\textbf{X} \neq \hat{\textbf{X}}} P\left(\textbf{X} \rightarrow \hat{\textbf{X}}\right)
\end{equation}

\noindent where $M$ denotes the signal constellation size and $k$ is the number of independent information symbols in the codeword. It is well known  \cite{TSC}, that an analysis of the PEP leads to the following design criteria:

$1)$. $Rank$ $criterion$: To achieve maximum diversity, the codeword difference matrix $(\textbf{X} - \hat{\textbf{X}})$ must be full rank for all possible pairs of codeword pairs and the diversity gain is given by $n_tn_r$. If full rank is not achievable, then, the diversity gain is given by $rn_r$, where $r$ is the minimum rank of the codeword difference matrix over all possible codeword pairs.

$2)$. $Determinant$ $criterion$: For a full ranked STBC, the minimum determinant $\delta_{min}$, defined as
\begin{equation}
 \delta_{min} = \min_{\textbf{X} \neq \hat{\textbf{X}}} det\left[\left(\textbf{X}-\hat{\textbf{X}}\right)\left(\textbf{X}-\hat{\textbf{X}}\right)^{H}\right]
\end{equation}
should be maximized. The coding gain is given by $\left(\delta_{min}\right)^{1/n_t}$, with $n_t$ being the number of transmit antennas.

If the STBC is non full-diversity and $r$ is the minimum rank of the codeword difference matrix over all possible codeword pairs, then , the coding gain $\delta$ is given by 
\begin{equation*}
\delta = \min_{\textbf{X} -\hat{\textbf{X}}}\left(\prod_{i=1}^{r}\lambda_i\right)^{\frac{1}{r}} 
\end{equation*}
\noindent where  $\lambda_i, i = 1,2,\cdots,r$, are the non-zero eigen values of the matrix $\left(\textbf{X} - \hat{\textbf{X}}\right)\left(\textbf{X} - \hat{\textbf{X}}\right)^H$ 

 It should be noted that, for high signal-to-noise ratio (SNR) values at each receive antenna, the dominant parameter is the diversity gain which defines the slope  of the CER curve. This implies that it is important to first ensure full diversity of the STBC and then try to maximize the coding gain.

For the $4\times2$ MIMO system, the objective is to design a code that is full-rate, i.e transmits 2 symbols per channel use,  has full diversity and allows simplified ML decoding.
\section{ THE PROPOSED STBC }
\label{sec3}
In this section,  we present our STBC for the $4\times 2$ MIMO system. The design is based on the CIOD for 4 antennas, whose structure is as defined below.
\begin{definition}
 CIOD for $4$ transmit antennas \cite{KhR} is as follows:\\
\noindent $\textbf{X}(s_1,s_2,s_3,s_4) =$
\begin{equation}
\label{ciod} 
\left[\begin{array}{cccc}
            s_{1I}+js_{3Q} & -s_{2I}+js_{4Q} & 0 & 0\\
            s_{2I}+js_{4Q} & s_{1I}-js_{3Q} & 0 & 0\\
            0 & 0 & s_{3I}+js_{1Q} & -s_{4I}+js_{2Q}\\
            0 & 0 & s_{4I}+js_{2Q} & s_{23}-js_{1Q}\\
           \end{array}\right]
\end{equation}
where $s_i \in \mathbb{C}, i = 1,\cdots,4$ are the information symbols and $s_{iI}$ and $s_{iQ}$ are the real and imaginary parts of $s_i$ respectively. Notice that in order to make the above STBC full rank, the signal constellation $\mathcal{A}$ from which the symbols are chosen should be such that the real part (imaginary part, resp.) of any signal point in $\mathcal{A}$ is not equal to the real part (imaginary part, resp.) of any other signal point in $\mathcal{A}$ \cite{KhR}. So if square or rectangular QAM constellations are chosen, they have to be rotated. The optimum angle of rotation, which we denote by $\theta_g$, has been found in \cite{KhR} to be $atan(2)/2$ degrees and this maximizes the diversity and coding gain.
\end{definition}

Our STBC is obtained as follows. Our $4\times4$ code matrix, denoted by $\textbf{S}$ encodes eight symbols $x_1,\cdots,x_8$ drawn from a QAM constellation, denoted by $\mathcal{A}_q$. We denote the rotated version of $\mathcal{A}_q$ by $\mathcal{A}$, with the angle of rotation chosen to be $\theta_g$ degrees. Let $s_i  \triangleq e^{j\theta_g}x_i, i = 1,2, \cdots 8$, so that the symbols $s_i$ are drawn from the constellation $\mathcal{A}$. The codeword matrix is defined as
\begin{equation}
 \textbf{S} \triangleq \textbf{X}(s_1,s_2,s_3,s_4) + e^{j\theta}\textbf{X}(s_5,s_6,s_7,s_8)\textbf{P}
\end{equation}
%
\noindent with $\theta \in [0,\pi/2]$ and $\textbf{P}$ being a permutation matrix designed to make the STBC full-rate, given by 
\begin{equation*}
 \textbf{P} = \left[\begin{array}{cccc}
0 & 0 & 1 & 0\\
0 & 0 & 0 & 1\\
1 & 0 & 0 & 0\\
0 & 1 & 0 & 0\\
\end{array}\right].
\end{equation*}
The choice of $\theta$ should be such that the diversity and coding gain are maximized. A computer search yielded the optimum value of $\theta$ to be $\pi/4$. This value of $\theta$ provides the largest coding gain achievable for this family of codes. The value of the minimum determinant obtained for unit energy 4-QAM constellation is 0.6400. The resulting code matrix is as shown in the top of the next page.
\begin{figure*}
\begin{eqnarray*}
 \textbf{S} = \left[\begin{array}{cccc}
            s_{1I}+js_{3Q} & -s_{2I}+js_{4Q} & e^{j\pi/4}(s_{5I}+js_{7Q}) & e^{j\pi/4}(-s_{6I}+js_{8Q})\\
            s_{2I}+js_{4Q} & s_{1I}-js_{3Q} & e^{j\pi/4}(s_{6I}+js_{8Q}) & e^{j\pi/4}(s_{5I}-js_{7Q})\\
            e^{j\pi/4}(s_{7I}+js_{5Q}) & e^{j\pi/4}(-s_{8I}+js_{6Q}) & s_{3I}+js_{1Q} & -s_{4I}+js_{2Q}\\
            e^{j\pi/4}(s_{8I}+js_{6Q}) & e^{j\pi/4}(s_{7I}-js_{5Q}) & s_{4I}+js_{2Q} & s_{3I}-js_{1Q}\\
           \end{array}\right]
\end{eqnarray*} \hrule
\end{figure*} 

\section{DECODING COMPLEXITY OF THE PROPOSED CODE}
\label{sec4}
The decoding complexity of the proposed code is of the order of $M^5$. This is due to the fact that conditionally given the symbols $x_5,x_6,x_7$ and $x_8$, the rest of the symbols $x_1,x_2,x_3,$ and $x_4$ can be decoded independent of one another. This can be shown as follows. Writing the STBC in terms of its linear weight matrices, we have
\begin{equation*}
 S  = \sum_{m=1}^{8}\underbrace{x_{mI}A_{2m-1} + x_{mQ}A_{2m}}_{\textbf{T}_m} =  S_1 + S_2
\end{equation*}
\noindent where
\begin{equation*}
 S_1 = \sum_{m=1}^{4}x_{mI}A_{2m} + x_{mQ}A_{2m+1}
\end{equation*}
\noindent and
\begin{equation*}
 S_2 = \sum_{m=5}^{8}x_{mI}A_{2m-1} + x_{mQ}A_{2m}.
\end{equation*}
The weight matrices are as follows
\begin{eqnarray*}
 A_{1} &  =  & \left[\begin{array}{rrrr}
                       cos\theta_g & 0 & 0 & 0 \\
                       0 & cos\theta_g & 0 & 0 \\
                       0 & 0 & jsin\theta_g & 0 \\
                       0 & 0 & 0 & -jsin\theta_g \\ 
                      \end{array}\right] \\
A_{2} &  =  & \left[\begin{array}{rrrr}
                       -sin\theta_g & 0 & 0 & 0 \\
                       0 & -sin\theta_g & 0 & 0 \\
                       0 & 0 & jcos\theta_g & 0 \\
                       0 & 0 & 0 & -jcos\theta_g \\ 
                      \end{array}\right] \\
A_{3} &  =  & \left[\begin{array}{rrrr}
                       0 & -cos\theta_g & 0 & 0 \\
                       cos\theta_g & 0 & 0 & 0 \\
                       0 & 0 & 0 & jsin\theta_g   \\
                       0 & 0 & jsin\theta_g & 0   \\ 
                      \end{array}\right] \\
A_{4} &  =  & \left[\begin{array}{rrrr}
                       0 & sin\theta_g & 0 & 0 \\
                       -sin\theta_g & 0 & 0 & 0 \\
                       0 & 0 & 0 & jcos\theta_g   \\
                       0 & 0 & jcos\theta_g & 0   \\ 
                      \end{array}\right] \\
A_{5} &  =  & \left[\begin{array}{rrrr}
                       jsin\theta_g & 0 & 0 & 0 \\
                       0 & -jsin\theta_g & 0 & 0 \\
                       0 & 0 & cos\theta_g & 0 \\
                       0 & 0 & 0 & cos\theta_g \\ 
                      \end{array}\right] \\
A_{6} &  =  & \left[\begin{array}{rrrr}
                       jcos\theta_g & 0 & 0 & 0 \\
                       0 & -jcos\theta_g & 0 & 0 \\
                       0 & 0 & -sin\theta_g & 0 \\
                       0 & 0 & 0 & -sin\theta_g \\ 
                      \end{array}\right] \\
A_{7} &  =  & \left[\begin{array}{rrrr}
                       0 & jsin\theta_g & 0 & 0 \\
                       jsin\theta_g & 0 & 0 & 0 \\
                       0 & 0 & 0 & -cos\theta_g   \\
                       0 & 0 & cos\theta_g & 0   \\ 
                      \end{array}\right] \\
A_{8} &  =  & \left[\begin{array}{rrrr}
                       0 & jcos\theta_g & 0 & 0 \\
                       jcos\theta_g & 0 & 0 & 0 \\
                       0 & 0 & 0 & sin\theta_g   \\
                       0 & 0 & -sin\theta_g & 0   \\ 
                      \end{array}\right] \\ 
A_{9} &  =  & e^{j\pi/4}\left[\begin{array}{rrrr}
                       0 & 0 & cos\theta_g & 0   \\
                       0 & 0 & 0 & cos\theta_g \\
                       sin\theta_g & 0 & 0 & 0  \\
                       0 & -sin\theta_g & 0 & 0 \\ 
                      \end{array}\right] \\
A_{10} &  =  & e^{j\pi/4}\left[\begin{array}{rrrr}
                       0 & 0 & -sin\theta_g & 0  \\
                       0 & 0 & 0 & -sin\theta_g   \\
                       cos\theta_g & 0 & 0 & 0 \\
                       0 & -cos\theta_g & 0 & 0 \\ 
                      \end{array}\right] \\
A_{11} &  =  & e^{j\pi/4}\left[\begin{array}{rrrr}
                       0 & 0 & 0 & -cos\theta_g   \\
                       0 & 0 & cos\theta_g & 0   \\
                       0 & sin\theta_g & 0 & 0    \\
                       sin\theta_g & 0 & 0 & 0   \\ 
                      \end{array}\right] 
\end{eqnarray*}

\begin{eqnarray*}
A_{12} &  =  & e^{j\pi/4}\left[\begin{array}{rrrr}
                       0 & 0 & 0 & sin\theta_g  \\
                       0 & 0 & -sin\theta_g & 0   \\
                       0 & cos\theta_g & 0 & 0   \\
                       cos\theta_g & 0 & 0 & 0  \\ 
                      \end{array}\right] \\
A_{13} &  =  & e^{j\pi/4}\left[\begin{array}{rrrr}
                       0 & 0 & sin\theta_g & 0   \\
                       0 & 0 & 0 & -sin\theta_g  \\
                       cos\theta_g & 0 & 0 & 0 \\
                       0 & cos\theta_g & 0 & 0 \\ 
                      \end{array}\right]\\ 
A_{14} &  =  & e^{j\pi/4}\left[\begin{array}{rrrr}
                       0 & 0 & cos\theta_g & 0   \\
                       0 & 0 & 0 & -cos\theta_g  \\
                       -sin\theta_g & 0 & 0 & 0 \\
                       0 & -sin\theta_g & 0 & 0 \\ 
                      \end{array}\right] \\
A_{15} &  =  & e^{j\pi/4}\left[\begin{array}{rrrr}
                       0 & 0 & 0 & sin\theta_g   \\
                       0 & 0 & sin\theta_g & 0 \\
                       0 & -cos\theta_g & 0 & 0    \\
                       cos\theta_g & 0 &  0 & 0  \\ 
                      \end{array}\right] \\
A_{16} &  =  & e^{j\pi/4}\left[\begin{array}{rrrr}
                       0 & 0 & 0 & cos\theta_g  \\
                       0 & 0 & cos\theta_g & 0   \\
                       0 & sin\theta_g & 0 & 0    \\
                       -sin\theta_g & 0 & 0 & 0   \\ 
                      \end{array}\right]\\
\end{eqnarray*}
\noindent Notice  that the matrix $S_1$ is as defined in \eqref{ciod}. The ML decoding metric in \eqref{ML} can be written as
\begin{eqnarray*}
M\left(S\right) & = & tr\left[\left(Y-HS\right)\left(Y-HS\right)^H\right]\\
& = & tr\left[\left(Y-HS_1-HS_2 \right)\left(Y-HS_1-HS_2 \right)^H\right]\\
& = & tr\left[\left(Y-HS_1\right)\left(Y-HS_1\right)^H\right] {}
                                                       \nonumber\\
&&{}-tr\left[HS_2 \left(Y-HS_1\right)^H\right]{}
                                          \nonumber\\
&&{}-tr\left[\left(Y-HS_1\right)\left(HS_2 \right)^H\right] {}
\nonumber\\
&&{}+ tr\left[HS_2 \left(HS_2 \right)^H\right]\\
\end{eqnarray*}
It can be verified that the following hold true for $l,m \in \left[1,8\right]$.
\begin{equation*}
 A_mA_l^H + A_lA_m^H = 0   \left\{ \begin{array}{ll}
\forall l \neq m, m+1, & \textrm{if } m \textrm{ is odd}\\
\forall l \neq m, m-1, & \textrm{if } m \textrm{ is even}\\
\end{array} \right.
\end{equation*}
\noindent From \cite{KhR}, we obtain
\begin{eqnarray*}
 tr\left[\left(Y-HS_1\right)\left(Y-HS_1\right)^H\right] = \\
 \sum_{m=1}^{4}\Vert Y - HT_m\Vert_F^2-3tr\left(YY^H\right)
\end{eqnarray*}
\noindent Therefore,
\begin{eqnarray*}
M\left(S\right) & = & \sum_{m=1}^{4}\Vert Y - HT_m\Vert_F^2-3tr\left(YY^H\right) {}
                                                                  \nonumber\\
&& {}+tr\left[HS_2 \left(HS_1\right)^H\right]+tr\left[HS_1\left(HS_2  \right)^H\right] {}
                                                                  \nonumber\\
&& {}-tr\left[HS_2  Y^H\right]-tr\left[Y\left(HS_2  \right)^H\right]{}
\nonumber\\
&&{}+tr\left[HS_2 \left(HS_2 \right)^H\right]
\end{eqnarray*}
\begin{eqnarray*}
& = & \sum_{m=1}^{4}\Vert Y - HT_m\Vert_F^2 +\sum_{m=1}^{4}tr\left[HS_2 \left(HT_m\right)^H\right]{} 
\nonumber\\
&& {}+\sum_{m=1}^{4}tr\left[HT_m\left(HS_2 \right)^H\right] +\Vert Y-HS_2 \Vert_F^2{}
\nonumber\\
&&{}-4tr(YY^H) 
\end{eqnarray*}
Hence, when $S_2 $ is given, i.e, symbols $x_5,x_6,x_7$ and $x_8$ are given, the ML metric can be decomposed as 
\begin{equation}
 M\left(S\right) = \sum_{m=1}^{4}M\left(x_m\right) + M_c
\end{equation}
\noindent with $M_c = \Vert Y-HS_2 \Vert_F^2{} -4tr(YY^H) $ and $M(s_m)$ being a function of symbol $x_m$ alone. Thus decoding can be done as follows: choose the quadruplet $\left(x_5,x_6,x_7,x_8\right)$ and then parallelly decode $x_1,x_2,x_3$ and $x_4$ so as to minimize the ML decoding metric. With this approach, there are $4M^5$ values of the decoding metric that need to be computed in the worst case. So, the decoding complexity is of the order of $M^5$.

\section{LOW COMPLEXITY DECODING USING SPHERE DECODER}
Now, we show how the sphere decoding can be used to achieve the decoding complexity of $M^5$.
It can be shown that \eqref{Y} can be written as 
\begin{equation}\label{eqmod}
 \widetilde{vec(\textbf{Y})} = \textbf{H}_{eq}\tilde{\textbf{s}} + \widetilde{vec(\textbf{N})}
\end{equation}
\noindent where $\textbf{H}_{eq} \in \mathbb{R}^{16\times16}$ is given by
\begin{equation}
 \textbf{H}_{eq} = \left(\textbf{I}_4 \otimes \check{\textbf{H}}\right)\textbf{G}
\end{equation}
with $\textbf{G} \in \mathbb{R}^{32\times16}$ being the generator matrix  for the STBC as defined in Definition \ref{def3} and 
\begin{equation*}
\tilde{\textbf{s}} \triangleq \left[\mathcal{R}(\textbf{s}_1),\mathcal{I}(\textbf{s}_1),\cdots,\mathcal{R}(\textbf{s}_8),\mathcal{I}(\textbf{s}_8)\right]^T.
\end{equation*}
\noindent with $s_i,i=1, \cdots, 8$, drawn from $\mathcal{A}$, which is a rotation of the regular QAM constellation $\mathcal{A}_q$. Let 
\begin{center} 
$\textbf{x}_q \triangleq [ x_1, x_2,\cdots,x_8 ]^T$ 
\end{center}
Then,
\begin{equation*}
 \tilde{\textbf{s}} = \textbf{F}\tilde{\textbf{x}}_q.
\end{equation*}
\noindent where $\textbf{F} \in \mathbb{R}^{16\times16}$ is $diag[\textbf{J},\textbf{J},\cdots,\textbf{J}]$ with $\textbf{J}$ being a rotation matrix and  defined as follows
\begin{equation*}
 \textbf{J} \triangleq \left[\begin{array}{cc}
                     cos(\theta_g) & -sin(\theta_g)\\
                     sin(\theta_g) &  cos(\theta_g)\\
              \end{array}\right]
\end{equation*}
So, \eqref{eqmod} can be written as 
\begin{equation}
\widetilde{vec(\textbf{Y})} = \textbf{H}_{eq}^\prime \tilde{\textbf{x}_q} + \widetilde{vec(\textbf{N})}
\end{equation}
\noindent where $\textbf{H}_{eq}^\prime = \textbf{H}_{eq}\textbf{F}$.
Using this equivalent model, the ML decoding metric can be written as
\begin{equation}
 \textbf{M}\left(\tilde{\textbf{x}_q}\right) = \Vert \widetilde{vec\left(\textbf{Y}\right)} - \textbf{H}_{eq}^\prime\tilde{\textbf{x}_q}\Vert^2.
\end{equation}
On obtaining the QR decomposition of $\textbf{H}_{eq}^\prime$, we get $\textbf{H}_{eq}^\prime $= $\textbf{QR}$, where  $\textbf{Q} \in \mathbb{R}^{16\times16}$ is an orthonormal matrix and  $\textbf{R} \in \mathbb{R}^{16\times16}$ is an upper triangular matrix. The ML decoding metric now can be written as
\begin{equation}
 \textbf{M}(\tilde{\textbf{x}_q}) = \Vert \textbf{Q}^T\widetilde{{vec(\textbf{Y})}} - \textbf{R}\tilde{\textbf{x}_q}\Vert^2.
\end{equation}
If $\textbf{H}_{eq}^\prime \triangleq [ \textbf{h}_1 \ \textbf{h}_2 \cdots \textbf{h}_{16} ]$, where $\textbf{h}_i, i = 1,2,\cdots,16$ are column vectors, then $\textbf{Q}$ and $\textbf{R}$ have the general form obtained by $Gram-Schmidt$ process as shown below
\begin{equation*}
 \textbf{Q} = [ \textbf{q}_1\ \textbf{q}_2 \ \textbf{q}_3 \cdots \textbf{q}_{16} ]
\end{equation*}
where $\textbf{q}_i, i = 1,2,\cdots,16$ are column vectors and 
\begin{equation*}
 \textbf{R} = \left[\begin{array}{ccccc}
\Vert \textbf{r}_1 \Vert & \langle \textbf{h}_2,\textbf{q}_1 \rangle & \langle \textbf{h}_3,\textbf{q}_1 \rangle & \ldots &  \langle \textbf{h}_{16},\textbf{q}_1 \rangle\\
0 & \Vert \textbf{r}_2 \Vert & \langle \textbf{h}_3,\textbf{q}_2 \rangle & \ldots & \langle \textbf{h}_{16},\textbf{q}_2 \rangle\\
0 & 0 &  \Vert \textbf{r}_3 \Vert & \ldots & \langle \textbf{h}_{16},\textbf{q}_3 \rangle\\ 
\vdots & \vdots & \vdots & \ddots & \vdots\\
0 & 0 & 0 & \ldots & \Vert \textbf{r}_{16} \Vert\\
\end{array}\right]
\end{equation*}
\noindent where\\
$\textbf{r}_1 = \textbf{h}_1$, $\ \textbf{q}_1 = \frac{\textbf{r}_1}{\Vert \textbf{r}_1 \Vert}$,\\
$\textbf{r}_i = \textbf{h}_i - \sum_{j=1}^{i-1}\langle \textbf{h}_i,\textbf{q}_j \rangle \textbf{q}_j$, $\ \textbf{q}_i = \frac{\textbf{r}_i}{\Vert \textbf{r}_i \Vert},\ i = 2,\cdots,16.$

It can be shown by direct computation that $\textbf{R}$ has the following structure
\begin{equation*}
 \textbf{R} = \left[ \begin{array}{cc}
                      \textbf{R}_1 & \textbf{R}_2\\
                      \textbf{O}_{8\times8} & \textbf{R}_3\\
              \end{array}\right]
\end{equation*}
\noindent where $\textbf{R}_1,\textbf{R}_2$ and $\textbf{R}_3$ are $8\times8$ matrices and $\textbf{R}_1$ specifically has the following structure
\begin{equation*}
 \textbf{R}_1 = \left[\begin{array}{rrrrrrrr}
                       a & a & 0 & 0 & 0 & 0 & 0 & 0 \\
                       0 & a & 0 & 0 & 0 & 0 & 0 & 0 \\
                       0 & 0 & a & a & 0 & 0 & 0 & 0 \\
                       0 & 0 & 0 & a & 0 & 0 & 0 & 0 \\
                       0 & 0 & 0 & 0 & a & a & 0 & 0 \\
                       0 & 0 & 0 & 0 & 0 & a & 0 & 0 \\
                       0 & 0 & 0 & 0 & 0 & 0 & a & a \\
                       0 & 0 & 0 & 0 & 0 & 0 & 0 & a \\
                      \end{array}\right]
\end{equation*}
\noindent and $\textbf{R}_3$, of course, is an upper triangular matrix.

The structure of the matrix $\textbf{R}$  allows us to perform an 8 dimensional real sphere decoding (SD) \cite{ViB} to find the partial vector $[ \mathcal{R}(x_5), \mathcal{I}(x_5),\cdots,\mathcal{I}(x_8) ]^T$ and hence obtain the symbols $x_5,x_6,x_7$ and $x_8$. Having found these, $x_1,x_2,x_3$ and $x_4$ can be decoded independently. Observe that the real and imaginary parts of symbol $x_1$ are entangled with one another because of constellation rotation but are independent of the real and imaginary parts of $x_2$, $x_3$ and $x_4$ when $x_5,x_6,x_7$ and $x_8$ are conditionally given. Similarly, $x_2$, $x_3$ and $x_4$ are independent of one another although their own real and imaginary parts are coupled with one another.

Having found the partial vector $[\mathcal{R}(x_5),\mathcal{I}(x_5),\cdots,\mathcal{I}(x_8) ]^T$, we proceed to find the rest of the symbols as follows. We do four parallel 2 dimensional real search to decode the symbols $x_1$, $x_2$, $x_3$ and $x_4$.  So, overall, the worst case decoding complexity of the proposed STBC is 4$M^5$. This due to the fact \\
1). An 8 dimensional real SD requires $M^4$ metric computations in the worst possible case.\\
2). Four parallel 2 dimensional real SD require $4M$ metric computations in the worst case.\\
This decoding complexity is significantly less than that for the BHV code proposed in \cite{BHV}, which is 2$M^7$ (as claimed in \cite{BHV}).

\section{SIMULATION RESULTS}
\label{sec5}
\begin{figure}
\centering
\includegraphics[width=3.7in,height=3.8in]{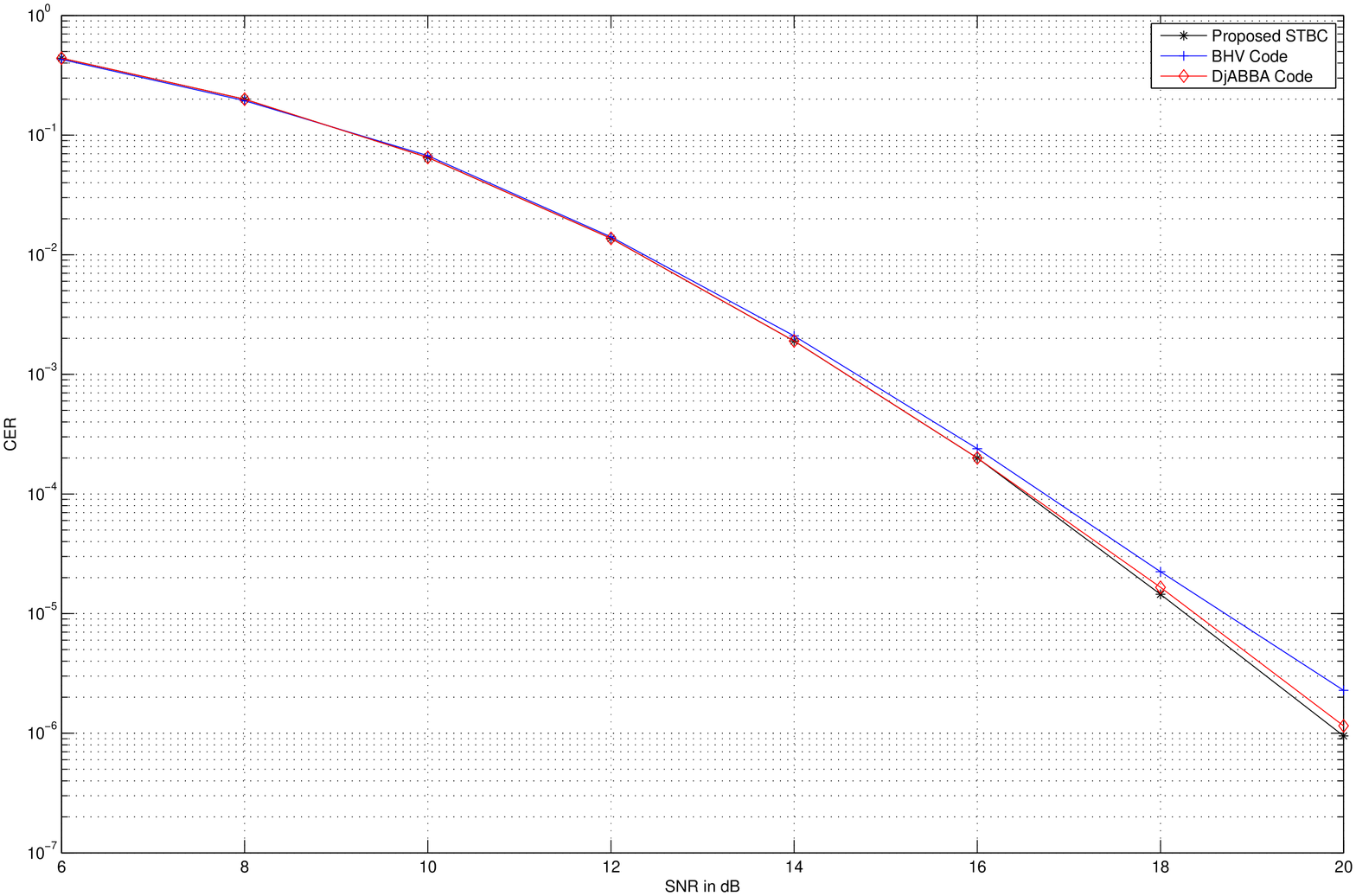}
\caption{CER performance for 4-QAM}
\label{4qam}
\end{figure}

\begin{figure}
\centering
\includegraphics[width=3.7in,height=3.8in]{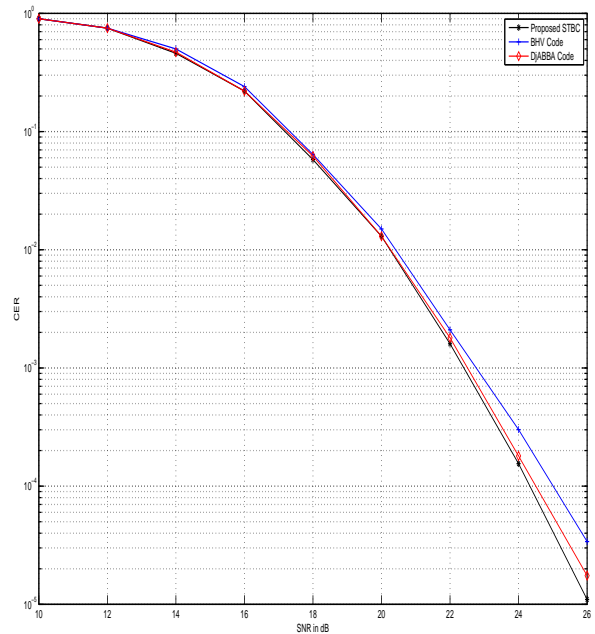}
\caption{CER performance for 16-QAM}
\label{16qam}
\end{figure}

We provide performance comparisons between the proposed code  and the  existing $4\times2$ full-rate codes - the DjABBA code \cite{HT}, \cite{HHVMM} and the BHV code. Fig \ref{4qam} shows the Codeword Error Rate (CER) performance plots for uncorrelated quasi-static Rayleigh flat-fading channel as a function of the received SNR at the receiver for 4-QAM signaling. All the codes perform similarly  at low and medium SNR. But at high SNR, the full diversity property of the DjABBA code and the proposed code  enables them to outperform the BHV code. In fact, our code slightly outperforms the DjABBA code at high SNR. Fig \ref{16qam} shows the CER performance for 16-QAM signaling, which shows a similar result. Table \ref{table1} gives a comparision of some of the well known codes for $4 \times 2$ MIMO systems

\begin{table*}
\begin{center}
\begin{tabular}{|c|c|c|c|c|} \hline
     & Min det  & \multicolumn{3}{c|}{ ML Decoding complexity}  \\ \cline{3-5}
 Code & for 4 QAM  & Square QAM &  Rectangular QAM & Non-rectangular  \\
      &          &            &  $M=M_1\times M_2$    &  QAM         \\ \hline
DjABBA code \cite{HT} &  0.04  & 4$M^6\sqrt{M}$ & $2M^6(M_1 + M_2)$ &  $2M^7$ \\ \hline
BHV code \cite{BHV} &  0  & 4$M^5$ & $2M^4(M_1^2 + M_2^2)$ &  $2M^6$ \\ \hline
The proposed code &  0.64 & $4M^5$ & $4M^5$  & $4M^5$ \\ \hline
\end{tabular}
\end{center}
\caption{COMPARISION BETWEEN THE MINIMUM DETERMINANT FOR 4-QAM AND DECODING COMPLEXITY OF SOME WELL KNOWN FULL-RATE $4\times2$ STBCs}
\label{table1}
\end{table*}


\section{DISCUSSION}
\label{sec6}
In this paper, we have presented a full-rate, full diversity  STBCs for $4\times2$ MIMO transmission which enables a significant reduction in the decoding complexity without having to pay in CER performance. In fact, our code  performs better than the  best known  full rate codes for $4\times2$ MIMO systems. So, to summarize, among the existing codes for $4$ transmit antennas and $2$ receive antennas, the proposed code is the best for QAM constellations .

\end{document}